\documentclass[12pt]{iopart}
\usepackage{graphicx,color,ulem}
\usepackage{amssymb,bm,iopams}

\renewcommand{\Re}{\mathop{\mathrm{Re}}}
\renewcommand{\Im}{\mathop{\mathrm{Im}}}

\begin{document}

\title[Spin squeezing in a generalized one-axis twisting model]{Spin
squeezing in a generalized one-axis twisting model}
\author{Guang-Ri Jin$^{1}$, Yong-Chun Liu$^{1}$, Wu-Ming Liu $^{2}$}

\address{$^1$ Department of Physics, Beijing Jiaotong
University, Beijing 100044, China}
\address{$^2$ Beijing National Laboratory for Condensed Matter Physics,
Institute of Physics, Chinese Academy of Sciences, Beijing 100080,
China} \ead{grjin@bjtu.edu.cn}

\begin{abstract}
We investigate the dependence of spin squeezing on the polar angle of the
initial coherent spin state $|\theta_0, \phi_0\rangle$ in a generalized
one-axis twisting model, where the detuning $\delta$ is taken into account.
We show explicitly that regardless of $\delta$ and $\phi_0$, previous
results of the ideal one-axis twisting is recovered as long as $%
\theta_0=\pi/2$. For a small departure of $\theta_0$ from $\pi/2$, however,
the achievable variance $(V_{-})_{\min}\sim N^{2/3}$, larger than the ideal
case $N^{1/3}$. We also find that the maximal-squeezing time $t_{\min}$
scales as $N^{-5/6}$. Analytic expressions of $(V_{-})_{\min}$ and $t_{\min}$
are presented, which agree with numerical simulations.
\end{abstract}

\maketitle

\section{Introduction}

Spin squeezing arising from quantum correlation of collective spin systems
\cite{Kitagawa}, has potential applications in high-precision measurement
\cite{Wineland1,Wineland2} and quantum information processes \cite%
{Sorensen,Pu,You,Wang,Wang2,Korbicz}. Kitagawa and Ueda have studied
the squeezing generated by a nonlinear Hamiltonian $\chi J_{z}^{2}$
due to the one-axis twisting (OAT) \cite{Kitagawa}. Starting from a
coherent spin state (CSS) \cite{CSS}: $|\theta _{0}=\pi /2,\phi
_{0}=0\rangle$, the system evolves into spin squeezed state (SSS),
which shows the reduced variance $(V_{-})$ below standard quantum
limit (SQL)---$N/4$, where $N$ is total particle number. The
smallest variance $(V_{-})_{\min}\sim N^{1/3}$ is obtainable at the
time scaled as $\chi t_{\min}\sim N^{-2/3}$.

Possible realization of the OAT-induced squeezing in a two-mode
Bose-Einstein Condensates (BECs) has been proposed \cite{Sorensen},
where the self-interaction parameter $\chi \sim
(a_{aa}+a_{bb}-2a_{ab})/2$ is inherently aroused from atomic intra-
and inter-species collisions. Atomic collisions lead to both the
squeezing and phase diffusion \cite{pd,Artur}. The dephasing process
destroys phase coherence of the two-component BECs, and thus sets a
limit to the applications of the condensates in high-precision
measurement and quantum information processing. A straightforward
way to suppress the diffusion is the preparation of number-squeezed
state, a special case of the SSS with the reduced variance along the
$J_{z}$ component. Such a kind of squeezed states have been
investigated both experimentally \cite%
{Orzel,Greiner,Strabley,Jaksch,Esteve,Chuu,Jo} and theoretically \cite%
{Bigelow,Law,Jin07,Jin08,Grond}.

Besides the above schemes that rely on nonlinear interactions of the
ultracold atoms, spin squeezing can be generated via light-matter
interactions \cite{Wineland1,Wineland2,Ueda,Saito,Hald} and quantum
nondemolition measurement \cite
{Kuzmich,Geremia,Wiseman,Jessen,Polzik,Takahashi,Takeuchi}.
Recently, the OAT-induced squeezing has been demonstrated in an
ensemble of cesium atoms \cite{Jessen,Polzik} and ytterbium atoms
\cite{Takahashi,Takeuchi}. In their experiments, the CSS with
$\theta_0=\pi/2$ was adopted as the input state, which is the
optimal initial state to obtain the strongest squeezing. Via optical
pumping, it was shown that 98\% atoms are in the CSS \cite{Polzik}.

In this paper, we investigate the degree of the OAT-induced
squeezing for $\theta_0$ slightly departure from $\pi/2$. A
generalized one-axis twisting model: $H=\delta J_{z}+\chi J_{z}^{2}$
is considered, which is the most important prototype in studying
spin squeezing \cite{Kitagawa,Sorensen} and quantum metrology
\cite{Boixo,Choi}. We prove explicitly that without particle losses,
the detuning $\delta$ and the azimuth angle $\phi_{0}$ give
vanishing contribution to the squeezing parameter, and the ideal
OAT-induced spin squeezing can be reproduced as long as
$\theta_{0}=\pi/2$. As the main result of our paper, we investigate
the dependence of the variance $(V_{-})_{\min }$ and the time
$t_{\min}$ on the particle number $N$ and the polar angle
$\theta_{0}$. Our results show that even for a small departure of
$\theta_{0}$ from $\pi/2$, power rule of the smallest variance
$(V_{-})_{\min}$ changes from $N^{1/3}$ to $N^{2/3}$ with the
increase of particle number $N$. The maximal squeezing is achievable
at the time that scaled as $\chi t_{\min}\sim N^{-5/6}$.

Our paper is organized as follows. In Sec. II, we present general formulas
of spin squeezing for arbitrary spin-1/2 system. In Sec. III, we study
quantum dynamics of the OAT model, which is exactly solvable for any initial
CSS. In Sec. IV, we present short-time solutions of the first- and
second-order moments of the spin operators. Approximated expression of the
reduced variance $V_{-}$ is presented to obtain power rules of the maximal
squeezing and its time scale $t_{\min}$. Finally, a summary of our paper is
presented.

\section{Some formulas of the spin squeezing}

\begin{figure}[tbp]
\centerline{
\includegraphics[width=5cm,angle=270]{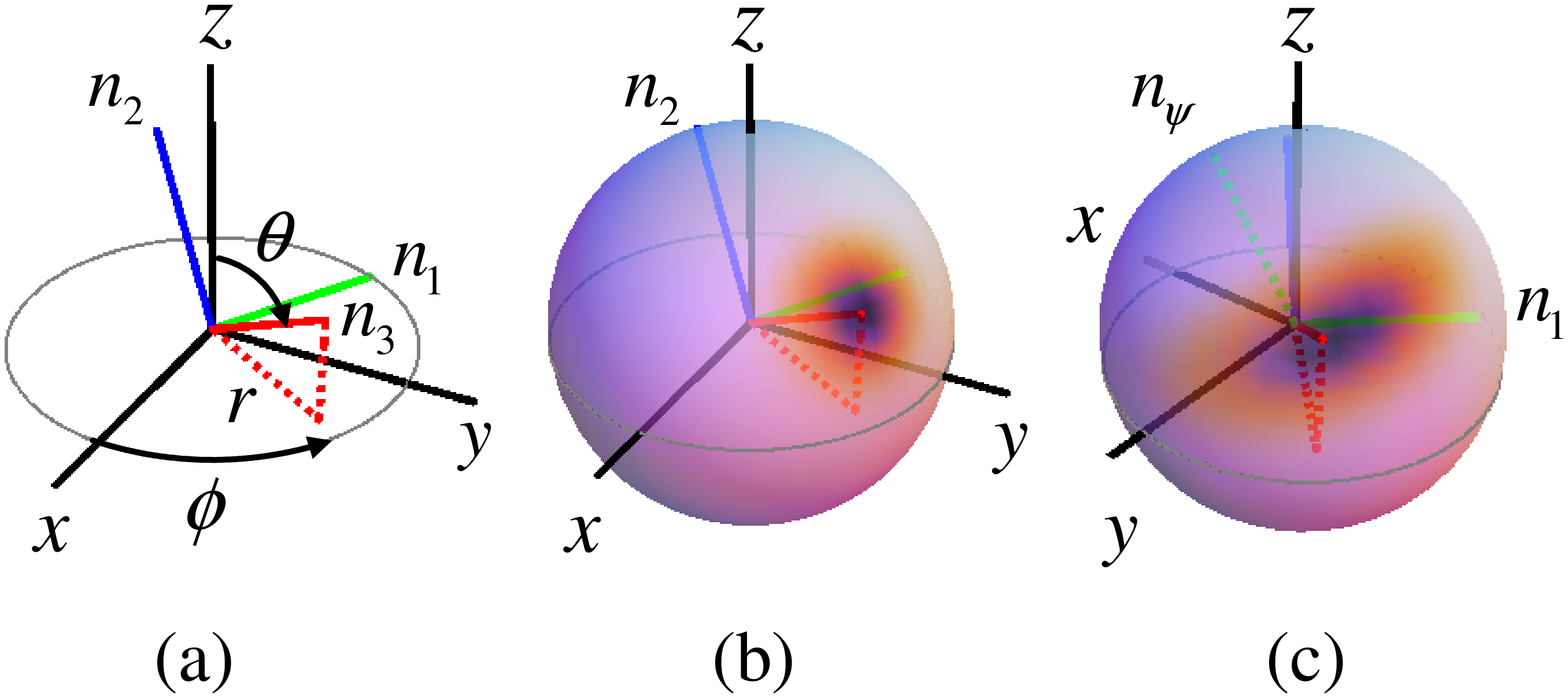}}
\caption{(color online) Husimi Q function: $Q(\protect\theta ,\protect\phi;
t)=|\langle \protect\theta, \protect\phi|\Psi(t)\rangle |^{2}$ on the Bloch
sphere for $j=30$. (a) the unit vectors $\mathbf{n}_1$ (green), $\mathbf{n}%
_2 $ (blue), $\mathbf{n}_3$ (red), as defined in Eq. (1). (b) the initial
CSS $|\protect\theta _{0}=\protect\pi /3, \protect\phi _{0}= \protect\pi %
/3\rangle$. (c) the SSS generated by the OAT Hamiltonian $H=\protect\chi %
J_z^2$ at time $t_{\min}=0.043\protect\chi ^{-1}$, where $\protect\chi %
t_{\min}$ is the time scale to attain the strongest squeezing. For large $j$%
, it is given by Eq. (\protect\ref{tmin}). }
\label{fig1}
\end{figure}

Assume that an ensemble of $N$ two-level atoms (i.e., spin $1/2$
particles) with ground state $|a\rangle $ and excited state
$|b\rangle $ can be described by collective spin operator
$\mathbf{J}=\sum_{k=1}^{N}\frac{1}{2}\mathbf{\sigma }^{(k)}$, where
$\mathbf{\sigma }^{(k)}$ is the Pauli operator of the $k$th atom.
Spin components of $\mathbf{J}$ obey SU(2) algebra,
$[J_{\mathbf{n}_{\mathbf{1}}},J_{\mathbf{n}_{\mathbf{2}}}]=iJ_{\mathbf{n}_{\mathbf{3}}}$
for any three orthogonal vectors $\mathbf{n}_{1}$, $\mathbf{n}_{2}$,
$\mathbf{n}_{3}$. The associated uncertainty relation reads $(\Delta
J_{\mathbf{n}_{\mathbf{1}}})^{2}(\Delta
J_{\mathbf{n}_{\mathbf{2}}})^{2}\geq \frac{1}{4}|\langle
J_{\mathbf{n}_{\mathbf{3}}}\rangle |^{2}$, where the variance is
defined as usual, $(\Delta \hat{A})^{2}=\langle \Psi
|\hat{A}^{2}|\Psi \rangle -\langle \Psi |\hat{A}|\Psi \rangle ^{2}$
for any spin state $|\Psi \rangle $ and
operator $\hat{A}$. Considering the mean spin $\langle \mathbf{J}%
\rangle =(\langle J_{x}\rangle ,\langle J_{y}\rangle ,\langle
J_{z}\rangle )$, we choose the orthogonal vectors as
\begin{eqnarray}
\mathbf{n}_{1} &=&\left( -\sin \phi ,\cos \phi ,0\right) ,  \nonumber \\
\mathbf{n}_{2} &=&\left( -\cos \theta \cos \phi ,-\cos \theta \sin \phi
,\sin \theta \right) ,  \label{n1n2n3} \\
\mathbf{n}_{3} &=&\left( \sin \theta \cos \phi ,\sin \theta \sin \phi ,\cos
\theta \right) ,  \nonumber
\end{eqnarray}%
where the azimuth angles $\phi =\tan ^{-1}[\langle J_{y}\rangle
/\langle J_{x}\rangle ]$, and the polar angle $\theta =\tan
^{-1}[r/\langle J_{z}\rangle ]$ with $r=|\langle J_{+}\rangle
|=(\langle J_{x}\rangle ^{2}+\langle J_{y}\rangle ^{2})^{1/2}$ [see
Fig.~\ref{fig1}(a)]. For arbitrary spin state $|\Psi \rangle $, it
is easy to prove that the mean spin $\langle \mathbf{J}\rangle $ is
along the $\mathbf{n}_{3}$ direction, with the length of the mean
spin $R=|\langle \mathbf{J}\rangle |=\langle
J_{\mathbf{n}_{\mathbf{3}}}\rangle $ [see Append.]. Now, let us
consider the CSS \cite{CSS}:
\begin{equation}
|\theta ,\phi \rangle =e^{-i\theta J_{\mathbf{n}_{\mathbf{1}}}}|j,j\rangle
=e^{i\theta (J_{x}\sin \phi -J_{y}\cos \phi )}|j,j\rangle ,  \label{CSS}
\end{equation}%
which is eigenstate of $J_{\mathbf{n}_{\mathbf{3}}}$ with eigenvalue
$j=N/2$ (where $N$ is total particle number), and thus $\langle
J_{\mathbf{n}_{\mathbf{3}}}\rangle =|\langle \mathbf{J}\rangle |=j$.
In single-particle picture, the CSS can be rewritten as a direct
product, $|\theta ,\phi \rangle =\prod_{k=1}^{N}[\cos (\theta /2)
|b\rangle _{k}+e^{i\phi }\sin (\theta /2)|a\rangle _{k}]$, where
$|a\rangle _{k} $ and $|b\rangle _{k}$ are ground and excited states
of the $k$th atom. Such a quantum uncorrelated state obeys the
minimal uncertainty relationship: $ (\Delta
J_{\mathbf{n}_{\mathbf{1}}})^{2}=(\Delta
J_{\mathbf{n}_{\mathbf{2}}})^{2}=\frac{1}{2}|\langle
J_{\mathbf{n}_{\mathbf{3}}}\rangle |=j/2$, where the value $j/2$ is
termed as the SQL.

Since the mean spin is parallel with $\mathbf{n}_{3}$, one can introduce any
spin component normal to the mean spin as
\begin{equation}
J_{\psi }=\mathbf{J}\cdot \mathbf{n}_{\psi }=J_{\mathbf{n}_{1}}\cos \psi +J_{%
\mathbf{n}_{2}}\sin \psi ,
\end{equation}%
where the unit vector $\mathbf{n}_{\psi }=\mathbf{n}_{1}\cos \psi +\mathbf{n}%
_{2}\sin \psi $, with $\psi $, being arbitrary angle between $\mathbf{n}_{1}$
and $\mathbf{n}_{\psi }$. For any spin state $|\Psi \rangle $, we have $%
\langle J_{\psi }\rangle =0$ and therefore, the variance of $J_{\psi}$ reads
\begin{equation}
(\Delta J_{\psi })^{2}=\frac{1}{2}\left[ \mathcal{C}+\mathcal{A}\cos (2\psi
)+\mathcal{B}\sin (2\psi )\right] ,  \label{variance}
\end{equation}%
where the coefficients $\mathcal{A}=\langle J_{\mathbf{n}_{\mathbf{1}%
}}^{2}-J_{\mathbf{n}_{\mathbf{2}}}^{2}\rangle $, $\mathcal{B}=\langle J_{%
\mathbf{n}_{\mathbf{1}}}J_{\mathbf{n}_{\mathbf{2}}}+J_{\mathbf{n}_{\mathbf{2}%
}}J_{\mathbf{n}_{\mathbf{1}}}\rangle $, and $\mathcal{C}=\langle J_{\mathbf{n%
}_{\mathbf{1}}}^{2}+J_{\mathbf{n}_{\mathbf{2}}}^{2}\rangle =j(j+1)-\langle
J_{\mathbf{n}_{\mathbf{3}}}^{2}\rangle $. Another orthogonal spin component
with respect to $J_{\psi }$ and its variance can be obtained by replacing $%
\psi $ with $\psi +\pi /2$. For the CSS $|\theta ,\phi \rangle $, it is easy
to verify that the coefficients $\mathcal{A}=\mathcal{B}=0$ and $\mathcal{C}%
=j$, which gives the variance $(\Delta J_{\psi })^{2}=j/2$, indicating
isotropically distributed variances of the CSS \cite{Kitagawa}, as shown in
Fig.~\ref{fig1}(b).

A spin-squeezed state (SSS) is defined if the variance of one spin component
normal to the mean spin is smaller than the SQL \cite{Kitagawa}, i.e., $%
(\Delta J_{\psi })^{2}<j/2$. The SSS has anisotropic variances distribution
in a plane normal the mean spin [see Fig.~\ref{fig1}(c)]. Optimally squeezed
angle $\psi_{op}$ is obtained via minimizing $(\Delta J_{\psi })^{2}$ with
respect to $\psi$, yielding $\tan (2\psi _{op})=\mathcal{B}/\mathcal{A}$, so
$\cos (2\psi_{op})=\pm \mathcal{A}/\sqrt{\mathcal{A}^{2}+\mathcal{B}^{2}}$
and $\sin (2\psi_{op})=\pm \mathcal{B}/\sqrt{\mathcal{A}^{2}+\mathcal{B}^{2}}
$. Substituting these results into Eq. (\ref{variance}), we obtain the
reduced and the increased variances \cite{Wang,Jin07,Jin08}
\begin{equation}
V_{\pm }=\frac{1}{2}\left[ \mathcal{C}\pm \sqrt{\mathcal{A}^{2}+\mathcal{B}%
^{2}}\right] ,  \label{V-}
\end{equation}%
where the reduced variance $V_{-}=(\Delta J_{\psi})^{2}$ corresponds to the
squeezing along $\mathbf{n}_{\psi}$ with $\psi=\psi_{op}=[\pi+\tan ^{-1}(%
\mathcal{B}/\mathcal{A})]/2$; while the increased variance $V_{+}$ gives the
so-called anti-squeezing for the angle $\psi=\psi_{op}+\pi/2$. The degree of
spin squeezing can be quantified by the normalized variance
\begin{equation}
\xi ^{2}=\frac{2V_{-}}{j}=\frac{\mathcal{C}-\sqrt{\mathcal{A}^{2}+\mathcal{B}%
^{2}}}{j}.  \label{xi}
\end{equation}%
For the CSS, the variances $V_{-}=V_{+}=j/2$ and $\xi ^{2}=1$; while for the
SSS, $\xi ^{2}<1$. It should be mentioned that the coefficients $\mathcal{A}$%
, $\mathcal{B}$, and $\mathcal{C}$ depend only on five quantities [see
Append. A]: $\langle J_{z}\rangle $, $\langle J_{+}\rangle $, $\langle
J_{z}^{2}\rangle $, $\langle J_{+}^{2}\rangle $, and $\langle
J_{+}(2J_{z}+1)\rangle $, from which one can solve the mean spin $\langle
\mathbf{J}\rangle $ and the squeezing parameter $\xi ^{2}$. In addition,\
there are several definitions of the squeezing parameter. According to
Wineland \textit{et al}. \cite{Wineland1}, the squeezing parameter is
defined as%
\begin{equation}
\zeta ^{2}=\frac{2j}{|\langle \mathbf{J}\rangle |^{2}}V_{-}=\frac{j^{2}}{%
|\langle \mathbf{J}\rangle |^{2}}\xi ^{2},  \label{xiR}
\end{equation}%
which closely relates to both frequency resolution in spectroscopy \cite%
{Wineland1} and many-body quantum entanglement \cite{Sorensen}.

\section{Generalized one-axis twisting model and its exact solutions}

The above formulas are valid for any spin-1/2 system with SU(2) symmetry. As
an example, we consider a two-component BECs \cite{Hall,BVHall} confined in
a deep 3D harmonic potential. The total system can be described by the
two-mode Hamiltonian ($\hbar =1$) \cite{Milburn}:
\begin{eqnarray}
H =\omega _{a}\hat{N}_{a}+\omega _{b}\hat{N}_{b}+U_{ab}\hat{N}_{a}\hat{N}%
_{b}+\frac{U_{aa}}{2}(\hat{a}^{\dag })^{2}(\hat{a})^{2}+\frac{U_{bb}}{2}(%
\hat{b}^{\dag })^{2}(\hat{b})^{2},  \label{H1}
\end{eqnarray}%
where $\hat{a}$, $\hat{b}$, and $\hat{N}_{i}$ ($i=a$, $b$) are the
annihilation and number operators for the two internal states $|a\rangle $
and $|b\rangle $, $\omega _{i}$ are single-particle kinetic energies, and $%
U_{ij}=(4\pi a_{ij}/M)\int d^{3}r|\Phi _{0}(r)|^{4}$ are atom-atom
interaction strengthes. For a conserved total particle number $N=\hat{N}_{a}+%
\hat{N}_{b}$, the two-mode model can be rewritten as $H=\delta J_{z}+\chi
J_{z}^{2}$, where the detuning $\delta =\omega _{b}-\omega
_{a}+(U_{bb}-U_{aa})(N-1)/2$, and $\chi =(U_{aa}+U_{bb}-2U_{ab})/2$. Angular
momentum operators $J_{+}=(J_{-})^{\dag }=\hat{b}^{\dag }\hat{a}$ and $%
J_{z}=(\hat{N}_{b}-\hat{N}_{a})/2$, satisfying SU(2) algebra.

Assumed that the two-mode system evolves from the CSS, $|\Psi (0)\rangle
=|\theta _{0},\phi _{0}\rangle =\sum_{m}c_{m}(0)|j,m\rangle $, with the
probability amplitudes \cite{CSS}
\begin{equation}
c_{m}=\sqrt{\frac{(2j)!}{(j+m)!(j-m)!}}\cos ^{j+m}\left( \frac{\theta _{0}}{2%
}\right) \sin ^{j-m}\left( \frac{\theta _{0}}{2}\right) e^{i(j-m)\phi _{0}},
\label{amplitude}
\end{equation}%
where the polar angles $\theta _{0}$ and $\phi _{0}$ determine population
imbalance and the relative phase between the two internal states \cite%
{Dutton,YunLi}. The state vector at any time $t$ reads%
\begin{equation}
|\Psi (t)\rangle =\sum_{m=-j}^{j}c_{m}e^{-i(\delta m+\chi m^{2})t}\left\vert
j,m\right\rangle ,  \label{psit}
\end{equation}%
where the self-interaction $\chi $ scrambles phase of each number state $%
|j,m\rangle $, and leads to spin squeezing \cite{Kitagawa,Sorensen} and
phase diffusion \cite{pd} of the two-mode BEC. In theory, the diffusion is
quantified by correlation function $\langle \hat{b}^{\dag }\hat{a}\rangle $
(i.e., $\langle J_{+}\rangle $), which decays exponentially with the time
scale $\chi t_{d}=j^{-1/2}$ for $\theta _{0}=\pi /2$. Such a kind of the
dephasing process has been observed in experiment by extracting the
visibility of the Ramsey fringe \cite{Artur}.

As an ideal case, spin squeezing induced by the OAT Hamiltonian $\chi
J_{z}^{2}$ has been investigated for the initial CSS $|\theta
_{0}=\pi/2,\phi _{0}=0\rangle $ \cite{Kitagawa}. For this special CSS, it
was shown the smallest variance $(V_{-})_{\min }\sim (2j)^{1/3}$ is
obtainable at the time $t_{\min }\sim (2j)^{-2/3}$. Based upon this, S{\o }%
rensen \textit{et al}. studied possible realization of the squeezing in $%
^{23}$Na atom BECs \cite{Sorensen}. More important, they proposed that the
squeezing parameter can be used as a probe of many-body entanglement. In
this paper, we investigate dynamical generation of the SSS in the
generalized OAT model from arbitrary CSS. We find that the power rules
change significantly even for $\theta_{0}\sim\pi/2$.

At first, we determine the mean spin $\langle \mathbf{J}\rangle =(\langle
J_{x}\rangle ,\langle J_{y}\rangle ,\langle J_{z}\rangle )$, where $\langle
J_{z}\rangle =j\cos (\theta _{0})$, $\langle J_{x}\rangle =\Re\langle
J_{+}\rangle $, and $\langle J_{y}\rangle =\Im\langle J_{+}\rangle $, with
\begin{eqnarray}
\left\langle J_{+}\right\rangle =j\sin \left( \theta _{0}\right) \exp \left[
i(\phi _{0}+\delta t)\right] \left[ \cos \left( \chi t\right) +i\cos \left(
\theta _{0}\right) \sin \left( \chi t\right) \right] ^{2j-1}.  \label{J+1}
\end{eqnarray}%
It is convenient to rewrite Eq. (\ref{J+1}) as $\langle J_{+}\rangle =r\exp
(i\phi )$, which yields $\langle J_{x}\rangle =r\cos \phi $ and $\langle
J_{y}\rangle =r\sin \phi $, as defined in Eq. (\ref{n1n2n3}). Therefore, we
obtain
\begin{eqnarray}
r &=&j\sin (\theta _{0})[1-\sin ^{2}\left( \theta _{0}\right) \sin
^{2}\left( \chi t\right) ]^{j-1/2},  \label{r} \\
\phi &=&\phi _{0}+\delta t+(2j-1)\varphi (t),  \label{phi}
\end{eqnarray}%
where $\varphi (t)=\tan ^{-1}[\cos (\theta _{0})\tan (\chi t)]$ is dynamical
phase. Note that in real calculations of the squeezing parameters, only $%
\cos (\phi )$ and $\sin (\phi )$ are needed and given by Eq. (\ref{cosphi})
and Eq. (\ref{sinphi}). The explicit form of the phase $\phi $ or $\varphi $
is introduced to find out the roles of $\delta $ and $\phi _{0}$ in the
squeezing. Obviously, $r$, $\varphi $, and also $R=(r^{2}+\langle
J_{z}\rangle ^{2})^{1/2}$ do not depend on them.

To proceed, we calculate the expectation values $\langle J_{z}^{2}\rangle$, $%
\langle J_{+}^{2}\rangle$, and $\langle J_{+}(2J_{z}+1)\rangle$, which are
relevant to the coefficients $\mathcal{A}$, $\mathcal{B}$, and $\mathcal{C}$%
. The mean value $\left\langle J_{z}^{2}\right\rangle $ reads
\begin{eqnarray}
\left\langle J_{z}^{2}\right\rangle =\frac{j}{2}\sin ^{2}\left( \theta
_{0}\right) +j^{2}\cos ^{2}\left( \theta _{0}\right) =\frac{j}{2}+j\left(
j-1/2\right) \cos ^{2}\left( \theta _{0}\right) ,  \label{Jz2}
\end{eqnarray}%
which, together with $\langle J_{z}\rangle =j\cos (\theta _{0})$, gives atom
number variance $(\Delta \hat{N}_{a})^{2}=(\Delta \hat{N}_{b})^{2}\equiv
(\Delta J_{z})^{2}=(j/2)\sin ^{2}\theta _{0}$. For $\theta _{0}\neq \pi /2$,
the variance $(\Delta J_{z})^{2}$ becomes narrow than that of the case $%
\theta _{0}=\pi /2$, which leads to relatively slow phase diffusion \cite%
{Sinatra,Jin09}. After some tedious calculations, we further obtain%
\begin{eqnarray}
\left\langle J_{+}^{2}\right\rangle &=&j\left( j-1/2\right) \sin ^{2}\left(
\theta _{0}\right) \exp \left[ 2i(\phi _{0}+\delta t)\right]  \nonumber \\
&&\times \left[ \cos \left( 2\chi t\right) +i\cos \left( \theta _{0}\right)
\sin \left( 2\chi t\right) \right] ^{2j-2},  \label{J+2}
\end{eqnarray}%
and
\begin{eqnarray}
\left\langle J_{+}(2J_{z}+1)\right\rangle &=&2j\left( j-1/2\right) \sin
\left( \theta _{0}\right) \exp \left[ i(\phi _{0}+\delta t)\right]  \nonumber
\\
&&\times \left[ \cos \left( \chi t\right) +i\cos \left( \theta _{0}\right)
\sin \left( \chi t\right) \right] ^{2j-2}  \nonumber \\
&&\times \left[ \cos \left( \theta _{0}\right) \cos \left( \chi t\right)
+i\sin \left( \chi t\right) \right] .  \label{J+2Jz+1}
\end{eqnarray}%
From Eq. (\ref{A})-Eq. (\ref{C}), one can find that the coefficients are
fully determined by five quantities: $\sin\theta$ ($=r/R$), $\cos\theta$ ($%
=\langle J_{z}\rangle /R$), $\langle J_{z}^{2}\rangle $, $\langle
J_{+}^{2}\rangle e^{-2i\phi }$, and $\langle J_{+}(2J_{z}+1)\rangle
e^{-i\phi }$. We have shown that the first three terms are independent with $%
\delta $ and $\phi _{0}$, which keeps true for the last two terms due to $%
\exp [ik(\phi _{0}+\delta t)]e^{-ik\phi }=\exp [-ik(2j-1)\varphi (t)]$ (with
$k=1,2$), where $\varphi (t)$ does not depend on $\delta $ and $\phi _{0}$.
As a result, we get the conclusion that the detuning $\delta $ and the
azimuth angle $\phi _{0}$ change the mean spin direction [see also Fig.~\ref%
{fig1}(b) and (c)], but do \textit{not} present any contribution to the
squeezing.

\begin{figure}[tbph]
\begin{center}
\includegraphics[width=12cm, angle=0]{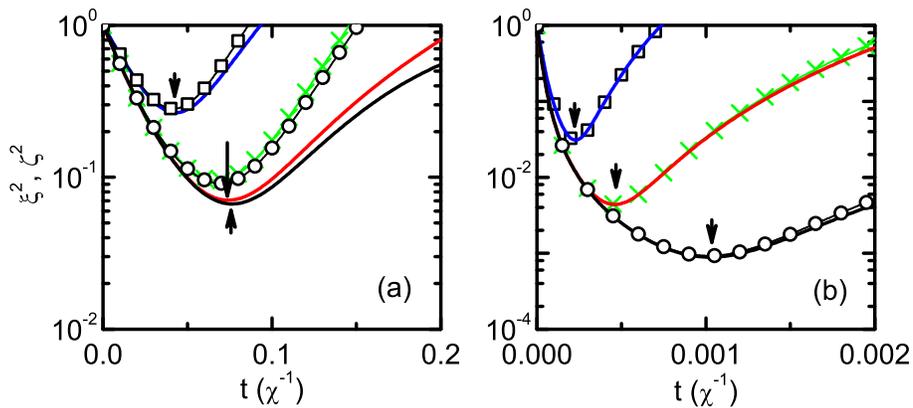}
\end{center}
\caption{(Color online) Time evolution of the squeezing parameters ($\protect%
\zeta ^{2}$, $\protect\xi ^{2}$) for various $\protect\theta _{0}$ of the
initial CSS. From top to bottom: $\protect\theta _{0}=\protect\pi /3$
(squares, blue lines), $0.98\times \protect\pi /2$ (crosses, red lines), and
$\protect\pi /2$ (empty circles, black lines). The arrows indicate the
positions of the maximal-squeezing time $t_{\min }$ for different $\protect%
\theta _{0}$'s. Other parameters: $j=30$ (a), $j=2\times 10^{4}$ (b), and $%
\protect\delta =\protect\phi _{0}=0$.}
\label{fig2}
\end{figure}

The squeezing parameters $\xi $ and $\zeta $ depend sensitively on the polar
angle $\theta _{0}$ of the initial CSS, as shown in Fig.~\ref{fig2}. The
most strongest squeezing can be obtained for $\theta _{0}=\pi /2$, which
corresponds to the initial CSS with equal atom population between the two
internal states, i.e., $\langle J_{z}\rangle =0$. From Eq. (\ref{J+1}), we
have $\langle J_{+}\rangle =r\exp (i\phi )$ with $r=R=j\cos ^{2j-1}(\chi t)$
and $\phi =\phi _{0}+\delta t$. From Eq. (\ref{Jz2})-Eq. (\ref{J+2Jz+1}), we
further obtain $\langle J_{z}^{2}\rangle =j/2$, $\langle J_{+}^{2}\rangle
e^{-2i\phi }=j(j-1/2)\cos ^{2j-2}(2\chi t)$, and $\langle
J_{+}(2J_{z}+1)\rangle e^{-i\phi }=ij(2j-1)\cos ^{2j-2}(\chi t)\sin (\chi t)$%
. Substituting these results into Eq. (\ref{A})-Eq. (\ref{C}), we obtain the
coefficients
\begin{eqnarray}
\mathcal{A} &=&\frac{j}{2}\left( j-1/2\right) \left[ 1-\cos ^{2j-2}(2\chi t)%
\right] , \\
\mathcal{B} &=&2j\left( j-1/2\right) \cos ^{2j-2}(\chi t)\sin (\chi t),
\end{eqnarray}%
and $\mathcal{C}=\mathcal{A}+j$. From Eq. (\ref{V-}), we get the increased
and the reduced variances%
\begin{equation}
V_{\pm }=\frac{j}{2}\left[ 1+\frac{j-1/2}{2}\left( \tilde{A}\pm \sqrt{\tilde{%
A}^{2}+\tilde{B}^{2}}\right) \right] ,  \label{vpi2}
\end{equation}%
where the intermediate coefficients $\tilde{A}=1-\cos ^{2j-2}(2\chi t)$ and $%
\tilde{B}=2\cos ^{2j-2}(\chi t)\sin (\chi t)$. One can find the variances
are exactly the same with that of ideal OAT case \cite{Kitagawa}, even for
nonzero $\delta $ and $\phi _{0}$.

Solid curves of Fig.~\ref{fig2} indicate the evolution of the normalized
variance $\xi ^{2}$. The minimal value of the squeezing parameter, $\xi
_{\min }^{2}=2j^{-1}(V_{-})_{\min }$, appears at the time $t_{\min }$
indicated by the arrows for different values of $\theta _{0}$. The smallest
value of $\xi _{\min }^{2}$ is obtained for the optimal initial state $%
\theta _{0}=\pi /2$. For $\theta _{0}\neq \pi /2$ and large $j$ ($>>1$), the
squeezing becomes worse than the optimal case. A closer look at the
evolution of $\zeta ^{2}$ [$=(j/|\langle \mathbf{J}\rangle |)^{2}\xi ^{2}$]
indicates that it is minimized before $t_{\min }$ [see empty circles of Fig.~%
\ref{fig2}(a)]. This is because different evolution rates of the variance $%
V_{-}$ and the mean spin $\langle \mathbf{J}\rangle $. In addition, the
minimal value $\zeta _{\min }^{2}$ is slightly larger than $\xi _{\min }^{2}$
due to the decreased mean spin $|\langle \mathbf{J}\rangle |\leq j$. For
large $j$ case, however, the two squeezing parameters almost merge with each
other in the short-time regime [see Fig.~\ref{fig2}(b)]. As a result, one
can assume that $\zeta _{\min }^{2}$ obeys the same power rule with $\xi
_{\min }^{2}$ \cite{Takeuchi}, and is determined by that of $(V_{-})_{\min }$%
.

\section{Power rules of the strongest squeezing and its time scale}

As shown by the red lines of Fig.~\ref{fig2}(b), both $\xi _{\min }^{2}$ and
$t_{\min }$ change significantly in comparison with the idea case (i.e., $%
\theta_0=\pi/2$). As a result, it is necessary to determine power rules of
the variance $(V_{-})_{\min }$ and the time $t_{\min }$ for $\theta
_{0}\neq\pi /2$. In this section, we calculate analytically the power rules
by using standard treatments of Ref. \cite{Kitagawa}. We will focus on a
small departure of $\theta _{0}$ from $\pi /2$ due to the fact that a
relatively small population imbalance between two internal states favors the
one-axis twisting effect.

\subsection{Ideal OAT case with $\protect\theta _{0}=\protect\pi /2$}

In the short-time limit ($\chi t<<1$) and large particle number ($j>>1$),
the increased and reduced variances Eq. (\ref{vpi2}) can be approximated as
\cite{Kitagawa}:
\begin{equation}
V_{+}\simeq \frac{j}{2}(4\alpha _{0}^{2}), V_{-}\simeq \frac{j}{2}\left(
\frac{1}{4\alpha _{0}^{2}}+\frac{2}{3}\beta _{0}^{2}\right) ,  \label{V+-}
\end{equation}%
where $\alpha _{0}=j\chi t>1$ and $\beta _{0}=j(\chi t)^{2}<<1$. Eq. (\ref%
{V+-}) is the key point to obtain the strongest squeezing $\xi _{\min }$ and
its time scale $t_{\min}$. Previously, the time $t_{\min }$ was obtained by
comparing the second term of $V_{-}$ with that of the first one \cite%
{Kitagawa}. Here, we solve $t_{\min }$ via minimizing $V_{-}$ with respect
to $t$, i.e.,
\begin{equation}
\left. \frac{d}{dt}(V_{-})\right\vert _{t_{\min }}=0,  \label{mini}
\end{equation}%
which yields power rule of the maximal-squeezing time:%
\begin{equation}
\chi t_{\min }\simeq 3^{1/6}(2j)^{-2/3}.  \label{t0}
\end{equation}%
Inserting $\chi t_{\min }$ into Eq. (\ref{V+-}), we further obtain the
reduced variance as
\begin{equation}
(V_{-})_{\min }\simeq \frac{3}{8}\left( \frac{2j}{3}\right) ^{1/3},
\label{V0}
\end{equation}%
and also, the smallest squeezing parameter $\xi _{\min
}^{2}=2j^{-1}(V_{-})_{\min }\simeq \frac{1}{2}(\frac{2j}{3})^{-2/3}$. Power
exponents of Eq. (\ref{t0}) and Eq. (\ref{V0}) are consistent with Ref. \cite%
{Kitagawa}, but different in the coefficients. As shown by the black solid
lines of Fig.~\ref{fig3}, the revised results fit very well with their
numerical results (empty circles).

\subsection{Small departure case with $\protect\theta _{0}\sim \protect\pi/2
$}

The power rules, Eq. (\ref{t0}) and Eq. (\ref{V0}), are valid only for $%
\theta _{0}=\pi /2$. Now, we generalize them for $\theta _{0}\neq \pi /2$
case. To obtain the approximated expressions of the variances as Eq. (\ref%
{V+-}), we calculate short-time solutions of $\langle J_{+}\rangle$, $%
\langle J_{+}^{2}\rangle$, and $\langle J_{+}(2J_{z}+1)\rangle$.

In the short-time limit ($\chi t<<1$), the dynamical phase $\varphi (t)=\tan
^{-1}[\cos (\theta _{0})\tan (\chi t)]\simeq \chi t\cos (\theta _{0})$, and
Eq. (\ref{J+1}) can be approximated as%
\begin{equation}
\left\langle J_{+}\right\rangle \simeq j\sin \left( \theta _{0}\right)
e^{i\phi }e^{-\beta },  \label{J+a}
\end{equation}%
where $\beta =\beta _{0}\sin ^{2}(\theta _{0})=j(\chi t)^{2}\sin ^{2}(\theta
_{0})$, and $\phi \simeq \phi _{0}+\delta t+2j\chi t\cos (\theta _{0})$. We
have assumed that particle number is large enough so $2j-1\simeq 2j$. The
length of the correlation reads $r=|\langle J_{+}\rangle |\simeq
j\sin(\theta _{0})e^{-\beta}$, which indicates that phase coherence of the
two-mode BEC decays exponentially (i.e., phase diffusion \cite{pd}) with the
coherence time scaled as $\chi t_{d}=\sin ^{-1}(\theta _{0})j^{-1/2}$ \cite%
{Sinatra,Jin09}. Similarly, short-time solutions of Eq. (\ref{J+2}) and Eq. (%
\ref{J+2Jz+1}) can be written approximately as
\begin{equation}
\left\langle J_{+}^{2}\right\rangle \simeq j\left( j-1/2\right) \sin
^{2}\left( \theta _{0}\right) e^{2i\phi }e^{-4\beta },  \label{J+2a}
\end{equation}%
and
\begin{equation}
\left\langle J_{+}(2J_{z}+1)\right\rangle \simeq j(2j-1)\sin \theta
_{0}(\cos \theta _{0}+i\chi t)e^{i\phi }e^{-\beta },  \label{J+2Jz+1a}
\end{equation}%
where the factor $\cos \theta _{0}$ can not be neglected since it is
comparable with $\chi t$. In fact, it is the presence of $\cos \theta _{0}$
that leads to significant change of $t_{\min }$ and $(V_{-})_{\min }$ even
for $\theta _{0}\sim \pi /2$.

\begin{figure}[tbph]
\begin{centering}
\includegraphics[width=12cm, angle=0]{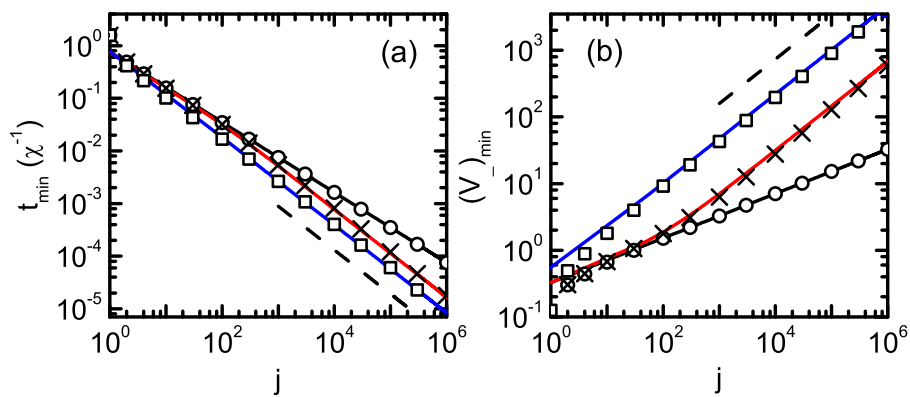}
\caption{(Color online) The maximal-squeezing time $t_{\min}$ (a),
and the
smallest variance $(V_{-})_{\min}$ (b) as a function of $j$ ($=N/2$) for $%
\protect\theta_0=\protect\pi/3$ (squares, blue lines), $0.98\times
\protect\pi/2$ (crosses, red lines), and $\protect\pi/2$ (empty
circles, black lines). Solid lines are predicted by Eq.
(\protect\ref{tmin}) and Eq. (\protect\ref{vmin}). Dashed lines are
plotted to guide the eyes, and are given by $\frac{1}{2}(2j)^{-5/6}$
(a) and $(2j)^{2/3}$ (b), respectively. Other parameters:
$\protect\delta=\protect\phi_0=0$, and $t_{\min}$ is in unit of
$\protect\chi^{-1}$.} \label{fig3}
\end{centering}
\end{figure}

To simplify the calculations, we make further approximations to the angles
of Eq. (\ref{A})-Eq. (\ref{C}): $\sin \theta=r/R\simeq \sin \theta _{0}$ and
$\cos \theta=\langle J_{z}\rangle/R\simeq \cos \theta _{0}$, where $%
\theta_{0}$ is polar angle of the initial CSS. This approximation is
equivalent with $r\simeq j\sin (\theta _{0})$, i.e., neglecting the the
diffusion within the squeezing time due to $t_{d}>t_{\min }$. Now, we expand
the coefficients $\mathcal{A}$, $\mathcal{B}$, and $\mathcal{C}$ in terms of
$\beta $. In calculating the increased variance, we only keep the lowest
order of $\beta $, and get $V_{+}\simeq \frac{j}{2}(4\alpha ^{2})$, where $%
\alpha =\alpha _{0}\sin ^{2}\theta _{0}=j\chi t\sin ^{2}\theta _{0}$. Next,
we solve power series of $4V_{+}V_{-}$ up to the third order of $\beta $,
from which we obtain the reduced variance as
\begin{equation}
V_{-}\simeq \frac{j}{2}\left[\frac{1}{4\alpha ^{2}}+\frac{2\beta ^{2}}{3}%
\left( 1+9j\sin ^{2}\theta _{0}\cos ^{2}\theta _{0}\right) \right] ,
\label{reduced V}
\end{equation}%
where the $j$-dependent additional term gives significant contribution to
the squeezing for $\theta_{0}\neq \pi /2$. By minimizing $V_{-}$ with
respect to $t$, we obtain power rule of the time as
\begin{equation}
\chi t_{\min }\simeq \frac{3^{1/6}(2j\sin ^{2}\theta _{0})^{-2/3}}{\left(
1+9j\sin ^{2}\theta _{0}\cos ^{2}\theta _{0}\right) ^{1/6}},  \label{tmin}
\end{equation}%
and that of the decreased variance:
\begin{equation}
(V_{-})_{\min }\simeq \frac{3}{8}\left[ \frac{2j}{3\sin ^{4}\theta _{0}}%
\left( 1+9j\sin ^{2}\theta _{0}\cos ^{2}\theta _{0}\right) \right] ^{1/3}.
\label{vmin}
\end{equation}%
For $\theta _{0}=\pi/2$, our results reduce to the ideal OAT case, i.e., Eq.
(\ref{t0}) and Eq. (\ref{V0}); while for $\theta _{0}\neq\pi/2$ and large $j$%
, Eq. (\ref{tmin}) and Eq. (\ref{vmin}) predict that the power rules change
to
\begin{equation}
\chi t_{\min }\sim (2j)^{-5/6}, (V_{-})_{\min }\sim (2j)^{2/3},
\end{equation}
which are confirmed by numerical simulations. To see this more clearly, let
us focus on red lines of Fig.~\ref{fig3}. For $\theta_0\sim\pi/2$ and small $%
j$, both the time $\chi t_{\min}$ and the variance $(V_{-})_{\min}$ follow
the same rule with the $\theta_0=\pi/2$ case. With the increase of $j$,
however, the red line (the crosses) of Fig.~\ref{fig3}(a) decreases faster
than the ideal OAT case [see also Fig.~\ref{fig2}(b)]. The change of the
power rule is shown more clearly in Fig.~\ref{fig3}(b).

\begin{figure}[tbph]
\begin{center}
\includegraphics[width=11cm, angle=0]{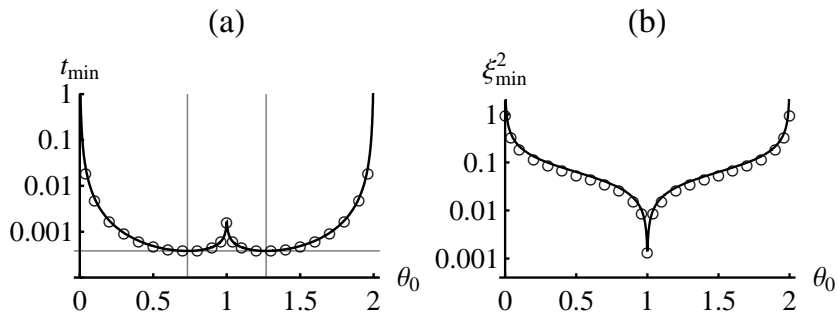}
\end{center}
\caption{The time $t_{\min }$ (a) and the normalized variance $\protect\xi %
_{\min }^{2}=2(V_{-})_{\min }/j$ (b) as a function of $\protect\theta _{0}$
(in unit of $\protect\pi /2$) for $j=10^{4}$. Empty circles are given by
numerical simulations, and solid lines are predicted by Eq. (\protect\ref%
{tmin}) and Eq. (\protect\ref{vmin}), respectively. Vertical grid lines in
(a) denote $\protect\theta _{0}=0.73\times \protect\pi /2$ and $\protect%
\theta _{0}=1.27\times \protect\pi /2$, separating monotonic regimes of $%
t_{\min }$. Other parameters are the same with Fig. 3.}
\label{fig4}
\end{figure}

In Fig.~\ref{fig4}, we show the dependence of $t_{\min }$ and
$(V_{-})_{\min }$ on $\theta _{0}$ for a fixed value $j$. It was
show that both $t_{\min }$ and $(V_{-})_{\min }$ are symmetrical
with respect to $\theta _{0}=\pi /2$.
The most strongest squeezing [i.e., the smallest value of $\xi _{\min }^{2}$%
] occurs for the optimal initial state $\theta _{0}=\pi /2$. Our analytic
results, Eq. (\ref{tmin}) and Eq. (\ref{vmin}), agree quite well with
numerical simulations except $\theta _{0}=0$ or $\pi $. In this case, the
state vector $|\Psi (t)\rangle =\exp [-i(\chi j^{2}\pm \delta j)t]|j,\pm
j\rangle $, which is the CSS with the variances $(V_{+})=(V_{-})=j/2$ and $%
\xi ^{2}=1$. However, Eq. (\ref{vmin}) diverges as $\theta _{0}\rightarrow 0$
or $\pi $, inconsistent with the real situation. Eq. (\ref{tmin}) gives
relatively good estimate of the maximal-squeezing time. As shown in Fig.~\ref%
{fig4}(a), $t_{\min }$ decreases monotonically in the small departure regime
$|\theta _{0}-\pi /2|<0.27\pi /2$, which implies that the maximal squeezing
occurs more and more earlier [see also Fig.~\ref{fig2}(b)]. Out of the
regime, $t_{\min }$ increases with the departure of $\theta _{0}$, and goes
infinity as $\theta _{0}\rightarrow 0$ or $\pi $.

\subsection{Dissipation effect due to atomic decay}

So far, we have neglected the effects of dissipation on the spin
squeezing, such as particle losses and center-of-motion of the atoms
in the BECs \cite{Sorensen,Dutton,YunLi}. For the squeezing
generated in atomic ensemble, the dominant dissipation source is
atomic decay due to spontaneous emission \cite{Polzik}, which can be
described by the master equation \cite{Saito}:
\begin{equation}
\frac{\partial \rho }{\partial t}=i[\rho ,H]+\frac{\gamma }{2}\left(
2J_{-}\rho J_{+}-J_{+}J_{-}\rho -\rho J_{+}J_{-}\right) ,
\end{equation}%
where $\rho $ is the density operator, and $\gamma$ is the decay
rate of the atoms. In the basis of $|j,m\rangle$, the elements $\rho
_{m,n}=\langle j,m|\rho |j,n\rangle$ could be solved numerically by
using the Runge-Kutta routine \cite{Saito}. In real calculations of
the squeezing parameters, only $6j$ elements like $\rho _{m,m}$,
$\rho _{m,m+1}$, and $\rho _{m,m+2}$ are needed.

\begin{figure}[tbph]
\begin{center}
\includegraphics[width=11cm, angle=0]{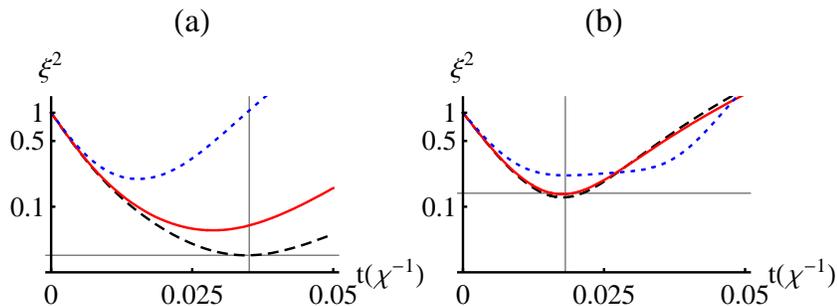}
\end{center}
\caption{(Color online) Time evolution of $\protect\xi^2$ for $\protect\theta%
_0=\protect\pi/2$ (a) and $\protect\theta_0=0.8\times\protect\pi/2$ (b).
From top to bottom, the decay rates $\protect\gamma=0.1\protect\chi$ (blue
dotted line), $\protect\gamma=0.01\protect\chi$ (red solid line), and $%
\protect\gamma=0$ (black dashed line). Vertical and horizontal grid
lines denote, respectively, $t_{\min}$ and $\protect\xi^2_{\min}$
for $\protect\gamma=0$ case. Other parameters: $j=100$,
$\protect\delta=\protect\phi_0=0$.} \label{fig5}
\end{figure}

In Fig. \ref{fig5}, we plot time evolution of $\xi ^{2}$ for small
decay rate, {\it e.g.}, $\gamma /\chi =0.01$ and $0.1$. Such a small
dissipation can be realized by increasing $\chi $, which in turn
leads to the preparation of the SSS within the lifetime of the atoms
$\gamma^{-1}$ \cite{Polzik}. For relatively small decay rate $\gamma
/\chi =0.01$ (red curves), both the maximal squeezing and its time
scale change slightly in comparison with $\gamma =0$ case. The
initial state with $\theta _{0}=\pi /2$ looks more sensitive to
atomic decay than $\theta _{0}=0.8\times \pi /2$ case. From the blue
dotted lines of Fig. \ref{fig5}(b), we find that even for $\gamma
/\chi =0.1$, a considerable squeezing with $\xi ^{2}$ (and also
$\zeta ^{2}$)$\sim 0.22$ could be reached in an ensemble of $200$
atoms, which occurs at a time scale given by Eq. (\ref{tmin}).

\section{Conclusion}

In summary, we have presented general formulas to study spin squeezing in
spin-1/2 system. Instead of six fluctuation parameters as Refs. \cite%
{Dutton,YunLi}, only five parameters, i.e., $\langle J_{z}\rangle $, $%
\langle J_{+}\rangle $, $\langle J_{z}^{2}\rangle $, $\langle
J_{+}^{2}\rangle $, and $\langle J_{+}(2J_{z}+1)\rangle $ are needed to
determine the mean spin and the squeezing parameters.

Spin squeezing of a generalized one-axis twisting model is investigated for
arbitrary coherent spin state $|\theta _{0},\phi _{0}\rangle $. We show
explicitly that $\theta _{0}=\pi /2$ is the optimal initial state to obtain
the minimum value of the variance $(V_{-})_{\min }\simeq \frac{3}{8}%
(2j/3)^{1/3}$, which takes place at the time scaled as $\chi t_{\min }\simeq
3^{1/6}(2j)^{-2/3}$. The detuning $\delta $ and the azimuth angle $\phi _{0}$
alert the mean spin's direction, but give vanishing contribution to the
squeezing parameters. As the main result of our paper, we calculate
analytically the dependence of the variance $(V_{-})_{\min }$ and the time $%
t_{\min }$ on the polar angle $\theta _{0}$, as Eq. (\ref{tmin}) and Eq. (%
\ref{vmin}), respectively. What may be a little surprising is that even for
a small departure of $\theta _{0}$ from $\pi /2$, the power rules become $%
(V_{-})_{\min }\sim (2j)^{2/3}$ and $\chi t_{\min }\sim (2j)^{-5/6}$,
deviating from the ideal case. The power rule, Eq. (\ref{tmin}%
), is robust against atomic decay for $\gamma <0.1\chi$. Our results
show that spin squeezed state in the OAT model depends sensitively
on the initial state and the interaction time. A straightforward way
to overcome these stringent requirements is still an open question.

\ack
G.R.J. would like to thank Professor C. K. Law for his invitation and
discussions at the Chinese University of Hong Kong. This work is supported
by the NSFC (Contract No.~10804007), the SRFDP (Contract No.~200800041003),
and Research Funds of Beijing Jiaotong University (Grants No.~2007XM049).
W.M.L. is supported by NSFC under Grants No. 60525417 and No. 10874235, and
the NKBRSFC under Grants No. 2006CB921400 and No. 2009CB930700.

\appendix

\section{The coefficients $\mathcal{A}$, $\mathcal{B}$, and $\mathcal{C}$}

In Eq. (\ref{n1n2n3}), we have defined three orthogonal unit vectors $%
\mathbf{n}_{i}$ ($i=1,2,3$), which are valid for any spin state $|\Psi
\rangle$. The angles $\theta$ and $\phi $ are determined by the mean spin $%
\langle \mathbf{J}\rangle =(\langle J_{x}\rangle ,\langle J_{y}\rangle
,\langle J_{z}\rangle )$, with
\begin{eqnarray}
\sin \theta &=&\frac{r}{R},\cos \theta =\frac{\langle J_{z}\rangle }{R},
\label{sin} \\
\cos \phi &=&\frac{\langle J_{x}\rangle }{r}=\frac{\Re\langle J_{+}\rangle }{%
r},  \label{cosphi} \\
\sin \phi &=&\frac{\langle J_{y}\rangle }{r}=\frac{\Im\langle J_{+}\rangle }{%
r},  \label{sinphi}
\end{eqnarray}%
where the length of the mean spin $R=|\langle \mathbf{J}\rangle |\equiv
(\langle J_{x}\rangle ^{2}+\langle J_{y}\rangle ^{2}+\langle J_{z}\rangle
^{2})^{1/2}$ and $r=|\langle J_{+}\rangle |\equiv (\langle J_{x}\rangle
^{2}+\langle J_{y}\rangle ^{2})^{1/2}=R\sin \theta $. From Eqs. (\ref{sin})-(%
\ref{sinphi}), it is easy to verify that the unit vector $\mathbf{n}%
_{3}=(\sin \theta \cos \phi ,\sin \theta \sin \phi ,\cos \theta)
=R^{-1}\langle \mathbf{J}\rangle $, i.e., the mean spin $\langle \mathbf{J}%
\rangle $ is parallel with the unit vector $\mathbf{n}_{3}$. Moreover, one
can prove the expectation value $\langle J_{\mathbf{n}_{\mathbf{1}}}\rangle
=-\langle J_{x}\rangle \sin \phi +\langle J_{y}\rangle \cos \phi =-\langle
J_{x}\rangle \langle J_{y}\rangle /r+\langle J_{y}\rangle \langle
J_{x}\rangle /r=0$, $\langle J_{\mathbf{n}_{\mathbf{2}}}\rangle =0$, and $%
\langle J_{\mathbf{n}_{\mathbf{3}}}\rangle =|\langle \mathbf{J}\rangle |$.

By using the above results, one can solve explicit expressions of the
coefficients: $\mathcal{A}=\langle J_{\mathbf{n}_{\mathbf{1}}}^{2}-J_{%
\mathbf{n}_{\mathbf{2}}}^{2}\rangle $, $\mathcal{B}=\langle J_{\mathbf{n}_{%
\mathbf{1}}}J_{\mathbf{n}_{\mathbf{2}}}+J_{\mathbf{n}_{\mathbf{2}}}J_{%
\mathbf{n}_{\mathbf{1}}}\rangle $, and $\mathcal{C}=\langle J_{\mathbf{n}_{%
\mathbf{1}}}^{2}+J_{\mathbf{n}_{\mathbf{2}}}^{2}\rangle $, yielding
\begin{eqnarray}
2\mathcal{A} &=&\sin ^{2}\theta \left[ j(j+1)-3\left\langle
J_{z}^{2}\right\rangle \right] -(1+\cos ^{2}\theta )\Re\left[ \left\langle
J_{+}^{2}\right\rangle e^{-2i\phi }\right]  \nonumber \\
&&+\sin (2\theta )\Re\left[ \left\langle J_{+}(2J_{z}+1)\right\rangle
e^{-i\phi }\right],  \label{A}
\end{eqnarray}%
\begin{eqnarray}
\mathcal{B}=-\cos (\theta )\Im\left[\left\langle
J_{+}^{2}\right\rangle e^{-2i\phi }\right] +\sin (\theta )\Im\left[
\left\langle J_{+}(2J_{z}+1)\right\rangle e^{-i\phi }\right],
\label{B}
\end{eqnarray}%
\begin{eqnarray}
\mathcal{C+A}=j(j+1)-\left\langle J_{z}^{2}\right\rangle -\Re\left[
\left\langle J_{+}^{2}\right\rangle e^{-2i\phi }\right] ,  \label{C}
\end{eqnarray}
where we have used the relations: $\langle J_{x}^{2}+J_{y}^{2}\rangle
=j(j+1)-\langle J_{z}^{2}\rangle $, $\langle J_{x}^{2}-J_{y}^{2}\rangle
=\Re\langle J_{+}^{2}\rangle $, $\langle J_{x}J_{y}+J_{y}J_{x}\rangle
=\Im\langle J_{+}^{2}\rangle $, $\langle J_{x}J_{z}+J_{z}J_{x}\rangle
=\Re\langle J_{+}(2J_{z}+1)\rangle $, and $\langle
J_{y}J_{z}+J_{z}J_{y}\rangle =\Im\langle J_{+}(2J_{z}+1)\rangle $.
Substituting the coefficients into Eq. (\ref{V-}), we obtain the variances $%
V_{\pm}$ and the squeezing parameters $\xi^{2}$ and $\zeta^{2}$.

Now let us calculate the coefficients for any CSS $|\theta ,\phi \rangle $.
The mean values $\langle J_{z}\rangle =j\cos (\theta )$ and $\left\langle
J_{+}\right\rangle =j\sin \left( \theta \right) e^{i\phi }$, which yields $%
\langle J_{x}\rangle =j\sin (\theta )\cos \phi $ and $\langle J_{y}\rangle
=j\sin (\theta )\sin \phi $. For the CSS, the length of the mean spin $R=j$.
These results can be directly obtained from Eq. (\ref{J+1}) by taking time $%
t=0$. Similarly, from Eqs. (\ref{Jz2})-(\ref{J+2Jz+1}), we further obtain $%
\langle J_{+}^{2}\rangle e^{-2i\phi}=j(j-1/2)\sin ^{2}(\theta)$ and $\langle
J_{+}(2J_{z}+1)\rangle e^{-i\phi}=j(2j-1)\sin (\theta)\cos(\theta)$, from
which we obtain immediately the coefficient $\mathcal{B}=0$. Substituting
these results into Eq. (\ref{A})-Eq. (\ref{C}), we also get $\mathcal{A}=0$,
and $\mathcal{C}=j$.


\end{document}